

\font\titlefont = cmr10 scaled\magstep 4
 2
\font\sectionfont = cmr10
\font\littlefont = cmr5 
\font\eightrm = cmr8

\def\ss{\scriptstyle}
\def\sss{\scriptscriptstyle}

\newcount\tcflag
\tcflag = 0  

\ifnum\tcflag = 0 \magnification = 1200 \fi  

\global\baselineskip = 1.2\baselineskip 
\global\parskip = 4pt plus 0.3pt 
\global\abovedisplayskip = 18pt plus3pt minus9pt
\global\belowdisplayskip = 18pt plus3pt minus9pt
\global\abovedisplayshortskip = 6pt plus3pt
\global\belowdisplayshortskip = 6pt plus3pt

\def\barsoff{\overfullrule=0pt}


\def\endignore{}
\def\ignore #1\endignore{} 

\newcount\dflag
\dflag = 0


\def\monthname{\ifcase\month 
\or January \or February \or March \or April \or May \or June%
\or July \or August \or September \or October \or November %
\or December 
\fi}

\newcount\dummy
\newcount\minute  
\newcount\hour
\newcount\localtime
\newcount\localday
\localtime = \time
\localday = \day

\def\advanceclock#1#2{ 
\dummy = #1
\multiply\dummy by 60
\advance\dummy by #2
\advance\localtime by \dummy
\ifnum\localtime > 1440 
\advance\localtime by -1440
\advance\localday by 1
\fi}

\def\settime{{\dummy = \localtime %
\divide\dummy by 60%
\hour = \dummy 
\minute = \localtime%
\multiply\dummy by 60%
\advance\minute by -\dummy 
\ifnum\minute < 10
\xdef\spacer{0} 
\else \xdef\spacer{}
\fi %
\ifnum\hour < 12
\xdef\ampm{a.m.} 
\else
\xdef\ampm{p.m.} 
\advance\hour by -12 %
\fi %
\ifnum\hour = 0 \hour = 12 \fi 
\xdef\timestring{\number\hour : \spacer \number\minute%
\thinspace \ampm}}}



\def\endtitle{}
\def\title#1\endtitle{\vskip.5in\titlefont
\global\baselineskip = 2\baselineskip 
#1\vskip.4in
\baselineskip = 0.5\baselineskip\rm}

\def\endauthors{}
\def\authors#1\endauthors{#1}

\def\endabstract{}
\def\abstract#1\endabstract{\vskip .3in%
\centerline{\sectionfont\bf Abstract}%
\vskip .1in
\noindent#1}

\def\nopageonenumber{\footline={\ifnum\pageno<2\hfil\else
\hss\tenrm\folio\hss\fi}}  

\newcount\nsection 
\newcount\nsubsection 

\def\section#1{\global\advance\nsection by 1
\nsubsection=0
\bigskip\noindent\centerline{\sectionfont \bf \number\nsection.\ #1}
\bigskip\rm\nobreak}

\def\subsection#1{\global\advance\nsubsection by 1
\bigskip\noindent\sectionfont \sl \number\nsection.\number\nsubsection)\
#1\bigskip\rm\nobreak}

\def\topic #1{{\medskip\noindent $\bullet$ \it #1:}}
\def\endtopic{\medskip}

\def\appendix#1#2{\bigskip\noindent%
\centerline{\sectionfont \bf Appendix #1.\ #2} 
\bigskip\rm\nobreak} 


\newcount\nref 
\global\nref = 1 

\def\therefs{}


\def\ref#1#2{\xdef #1{[\number\nref]} 
\ifnum\nref = 1\global\xdef\therefs{\item{[\number\nref]} #2\ } 
\else
\global\xdef\oldrefs{\therefs}
\global\xdef\therefs{\oldrefs\vskip.1in\item{[\number\nref]} #2\ }%
\fi%
\global\advance\nref by 1
}

\def\listrefs{\vfill\eject\section{References}\therefs}


\newcount\nfoot 
\global\nfoot = 1 

\def\foot#1#2{\xdef #1{(\number\nfoot)} 
\hskip -0.2cm ${}^{\number\nfoot}$
\footnote{}{\vbox{\baselineskip=10pt
\eightrm \hskip -1cm ${}^{\number\nfoot}$ #2}}
\global\advance\nfoot by 1
}


\newcount\nfig 
\global\nfig = 1
\def\thefigs{} 

\def\figure#1#2{\xdef #1{(\number\nfig)}
\ifnum\nfig = 1\global\xdef\thefigs{\item{(\number\nfig)} #2\ }
\else
\global\xdef\oldfigs{\thefigs}
\global\xdef\thefigs{\oldfigs\vskip.1in\item{(\number\nfig)} #2\ }%
\fi%
\global\advance\nfig by 1 } 

\def\fig#1{\xdef #1{(\number\nfig)}
\global\advance\nfig by 1 } 


\newcount\ntab
\global\ntab = 1

\def\table#1{\xdef #1{\number\ntab}
\global\advance\ntab by 1 } 


\newcount\cflag
\newcount\nequation
\global\nequation = 1
\def\eqlabel{(1)}

\def\nexteqno{\ifnum\cflag = 0
\global\advance\nequation by 1
\fi
\global\cflag = 0
\xdef\eqlabel{(\number\nequation)}}

\def\lasteqno{\global\advance\nequation by -1
\xdef\eqlabel{(\number\nequation)}}

\def\label#1{\xdef #1{(\number\nequation)}
\ifnum\dflag = 1
{\escapechar = -1
\xdef\draftname{\littlefont\string#1}}
\fi}

\def\clabel#1#2{\xdef\eqlabel{(\number\nequation #2)}
\global\cflag = 1
\xdef #1{\eqlabel}
\ifnum\dflag = 1
{\escapechar = -1
\xdef\draftname{\string#1}}
\fi}

\def\cclabel#1#2{\xdef\eqlabel{#2)}
\global\cflag = 1
\xdef #1{\eqlabel}
\ifnum\dflag = 1
{\escapechar = -1
\xdef\draftname{\string#1}}
\fi}


\def\eeq{}

\def\eqnn #1\eeq{$$ #1 $$}

\def\eq #1\eeq{
\ifnum\dflag = 0
{\xdef\draftname{\ }}
\fi 
$$ #1
\eqno{\eqlabel \rlap{\ \draftname}} $$
\nexteqno}







\def\eqa #1\eeq{
\ifnum\dflag = 0
{\xdef\draftname{\ }}
\fi 
$$ \eqalignno{ #1 } $$
\global\cflag = 0}


\def\ie{{\it i.e.\/}}


\def\plb#1#2#3{{\it Phys.\ Lett.} {\bf #1B} (19#2) #3}

\def\prd#1#2#3{{\it Phys.\ Rev.} {\bf D#1} (19#2) #3}


\global\nulldelimiterspace = 0pt



\def\frac#1#2{{{#1} \over {#2}}\,}  
\def\hf{{1\over 2}}



\def\Asl{\hbox{/\kern-.7500em\it A}} 
\def\Dsl{\hbox{/\kern-.6700em\it D}} 
\def\dsl{\hbox{/\kern-.5300em$\partial$}}
\def\pxpsl{\hbox{/\kern-.5600em$p$}}
\def\sslsh{\hbox{/\kern-.5300em$s$}}
\def\epssl{\hbox{/\kern-.5100em$\epsilon$}}
\def\delsl{\hbox{/\kern-.6300em$\nabla$}}
\def\lxpsl{\hbox{/\kern-.4300em$l$}}
\def\elxpsl{\hbox{/\kern-.4500em$\ell$}}
\def\kxpsl{\hbox{/\kern-.5100em$k$}}
\def\qxpsl{\hbox{/\kern-.5000em$q$}}
\def\sla#1{\raise.15ex\hbox{$/$}\kern-.57em #1}



\def\roughly#1{\mathrel{\raise.3ex\hbox{$#1$\kern-.75em\lower1ex\hbox{$\sim$}}}}

\def\gsim{\roughly>}



\def\bfk{{\bf k}}

\def\bfp{{\bf p}}


\def\Bfa{{\bf A}}
\def\Bfb{{\bf B}}

\def\Bfd{{\bf D}}
\def\Bfe{{\bf E}}


\def\Scl{{\cal L}}


\def\sst{{\sss T}}


\def\pmb#1{\setbox0=\hbox{#1}%
\kern-.025em\copy0\kern-\wd0
\kern.05em\copy0\kern-\wd0
\kern-.025em\raise.0433em\box0}


\font\jlgtenbrm=cmbx10
\font\jlgtenbit=cmmib10
\font\jlgtenbsy=cmbsy10
\font\jlgsevenbrm=cmbx10 at 7pt
\font\jlgsevenbsy=cmbsy10 at 7pt
\font\jlgsevenbit=cmmib10 at 7pt
\font\jlgfivebrm=cmbx10 at 5pt
\font\jlgfivebsy=cmbsy10 at 5pt
\font\jlgfivebit=cmmib10 at 5pt
\newfam\jlgbrm

\textfont\jlgbrm=\jlgtenbrm
\scriptfont\jlgbrm=\jlgsevenbrm
\scriptscriptfont\jlgbrm=\jlgfivebrm
\newfam\jlgbit

\textfont\jlgbit=\jlgtenbit
\scriptfont\jlgbit=\jlgsevenbit
\scriptscriptfont\jlgbit=\jlgfivebit
\newfam\jlgbsy

\textfont\jlgbsy=\jlgtenbsy
\scriptfont\jlgbsy=\jlgsevenbsy
\scriptscriptfont\jlgbsy=\jlgfivebsy
\newcount\jlgcode
\newcount\jlgfam
\newcount\jlgchar
\newcount\jlgtmp
\def\bolded#1{
        \jlgcode\the#1 \divide\jlgcode by 4096
        \jlgtmp\the\jlgcode \multiply\jlgtmp by 4096
        \jlgfam\the#1 \advance\jlgfam by -\the\jlgtmp
        \divide\jlgfam by 256
        \jlgtmp\the\jlgcode \multiply\jlgtmp by 16
	\advance\jlgtmp by \the\jlgfam
	\multiply\jlgtmp by 256
        \jlgchar\the#1 \advance\jlgchar by -\the\jlgtmp
        \advance\jlgfam by \the\jlgbrm
        \jlgtmp\the\jlgcode
        \multiply\jlgtmp by 16
        \advance\jlgtmp by \the\jlgfam
        \multiply\jlgtmp by 256
        \advance\jlgtmp by \the\jlgchar
        \mathchar\the\jlgtmp
}









\input epsf.tex


\nopageonenumber
\baselineskip = 18pt
\barsoff

\def\bk{\item{}}


\line{hep-th/9706449 \hfil McGill-97/10}

\vskip .1in
\title
\centerline{NRQED and Next-to-Leading}
\centerline{Hyperfine Splitting in Positronium}
\endtitle

\vskip 0.5in
\authors
\centerline{P. Labelle, S.M. Zebarjad and C.P. Burgess}
\vskip .2in
\centerline{\it  Physics Department, McGill University}
\centerline{\it 3600 University Street, Montr\'eal, Qu\'ebec, Canada, H3A 2T8.}
\endauthors

\abstract
\vbox{\baselineskip 15pt
We compute the next-to-leading, $O(m \alpha^5)$, contribution to the
hyperfine splitting in positronium within the framework of
NRQED. When applied to the ground state, our calculation
reproduces known results, providing a further test of NRQED
techniques. Besides providing a very simple method of
calculation of the standard result, we also obtain new
expressions for excited states of positronium with
negligible additional effort. Our calculation requires the
complete next-to-leading matching of the lowest-dimension
NRQED four-fermi couplings, which we publish here for
the first time.}
\endabstract


\vfill\eject

\section{Introduction and Summary}

\ref\lepage{W.E. Caswell and G.P. Lepage, \plb{167}{86}{437};\bk
T. Kinoshita and G.P. Lepage, in {\it Quantum Electrodynamics},
ed. by T. Kinoshita, (World Scientific, Singapore, 1990), pp. 81--89.}

The detailed comparisons between the heroic precision calculations
 using Quantum Electrodynamics (QED) and equally heroic
precision measurements  represent one of the pinnacles of
twentieth-century physics. On the theoretical side, comparisons
involving the energy levels of bound states, such as hydrogen or
positronium, pose a particular challenge. This is because of the
necessity of incorporating the small radiative corrections of
relativistic QED into the nonperturbative treatment required
for the bound state itself.
Important conceptual progress in handling bound states
in QED was made several years ago \lepage\ with the development
of nonrelativistic QED (NRQED), which consists
of an application of effective-field-theory ideas to atomic physics
applications of QED.

NRQED starts with the recognition that much of the complication
of QED bound-state calculations arises because the fine-structure
constant,\foot\units{We use throughout units for which $\ss
\hbar = c = 1$.} $\alpha = e^2/4 \pi$, enters into observables
in two conceptually different ways. First, $\alpha$ enters
as the small parameter which controls the higher-order QED
radiative corrections. Second, $\alpha$ enters from
the appearance in these calculations of three separate scales:
the electron mass, $m$, the bound-state momentum, $m v$,
and the bound-state energy, $m v^2$, once it is recognized
that $v = O(\alpha)$ in the bound state. While much of the
higher-order QED radiative corrections involve scales at, or above,
$m$, where relativistic kinematics is important, the complications
associated with handling the bound states all arise at the lower
two scales, $mv$ and $mv^2$.

NRQED takes advantage of this hierarchy of scales to efficiently
separate the radiative corrections from the bound-state physics.
First one accurately integrates out all physics associated with
momenta $p \gsim O(m)$, obtaining an effective theory of
nonrelativistic particles whose interactions are organized according
to their suppression by powers of the two independent small parameters,
$\alpha$ and $v$. This effective theory is then used to systematically compute
bound-state properties, at which point $v$ becomes of order $\alpha$.
Keeping $v$ and $\alpha$ independent until this last step makes the
bookkeeping more straightforward. The additional bonus is that
bound state calculations are much easier to do within the
effective theory because its nonrelativistic framework
permits the direct application of well-tested techniques based on
Schr\"odinger's equation.

\ref\patrick{P. Labelle, {\it Effective Field Theories for QED Bound
States: Extending Nonrelativistic QED to Study Retardation
Effects}, preprint McGill-96/33 (hep-ph/9608491).}

\ref\lambshift{P. Labelle and S.M. Zebarjad, {\it Derivation of the
Lamb Shift Using an Effective Field Theory}, preprint McGill-96/41
(hep-ph/9611313).}

Any effective field theory relies on the existence of a hierarchy of
scales, say $M_1 \ll M_2$,  in a physical problem. The heart of the
effective theory's utility lies in its powercounting rules, which
identify how to systematically isolate the complete
contributions to any observable to any fixed order in the small
ratio, $M_1/M_2$. NRQED is no exception in this regard, with the
powercounting rules identifying the suppression of observables
in powers of $v \sim \alpha$. It is the recent development of
NRQED power-counting rules \patrick, which now makes it
possible to directly identify the $O(\alpha^n)$ contributions to
any atomic-physics observable. These rules have recently been
demonstrated in practice in an illustrative calculation of the
Lamb shift \lambshift.

\ref\scalar{P. Labelle and S.M. Zebarjad, {\it An Effective Field Theory
Approach to Nonrelativistic Bound States in QED and Scalar QED},
preprint McGill-97/01.}

\ref\alphasixth{A. Hoang, P. Labelle and
 S.M. Zebarjad, work
in preparation.}

Although contributions order by order in $\alpha$
can also be obtained by other methods,
the virtue of NRQED lies in its simplicity, since this potentially
brings more complicated calculations into the domain
of the feasible. Besides permitting the very simple
extension of low-order results to bound states involving
particles of other spins \scalar, NRQED's simplicity
has very recently been exploited to give the first-ever
analytical calculation of the last previously uncomputed
$O(m \alpha^6)$ contribution to
positronium hyperfine splitting \alphasixth.

\ref\theliterature{R. Karplus  and  A. Klein, {\it Phys. Rev.}
{\bf 87} (1952) 848.}
\ref\swave{S.Gupta, W. Repko and C. Suchyta III, {\it Phys. Rev}
{\bf D40} (1989) 4100 .}

In the present paper we continue this process in other directions.
Since the power of effective field theory methods are best seen
once one goes beyond leading order in small parameters, we
present here an NRQED calculation of the complete contributions to the
positronium hyperfine structure at next-to-leading order, $O(m \alpha^5)$.
The result we obtain for the ground state
hyperfine splitting agrees with previous results \theliterature.
However our results are easily extended to the hyperfine splitting
of excited states having arbitrary quantum numbers
$n$ and $\ell$. These agree with previous results \swave\ for arbitrary $n$ but
$\ell = 0$, but to our knowledge ours is the first calculation which
is applicable to general $n$ and $\ell$.

Our goal in presenting this calculation is twofold. One of our
aims is to make NRQED accessible to a wider community. Detailed
calculations which can be compared to known results partly act
as a vehicle for explaining how NRQED works. On the other hand,
some of our results are new and interesting in their own right.
Our new results come in two forms. First, an intermediate step
in our calculation is the matching from QED to NRQED at
next-to-leading order in $\alpha$. Only part of this matching is
already available in the literature, and the remainder, which we publish
here for the first time, will have applications to many other higher-precision
NRQED calculations. In addition, our results for the hyperfine structure
of the excited states of positronium are also new.

\ref\thesis{P. Labelle, {\it Order $\alpha^2$ Nonrelativistic
Corrections to Positronium Hyperfine Splitting and Decay Rate
Using Nonrelativistic Quantum Electrodynamics}, Cornell
University Ph.D. thesis, UMI-94-16736-mc (microfiche), Jan 1994.}

We organize our presentation as follows. In \S2 we briefly review
the NRQED lagrangian, including a discussion of the matching
which is required to obtain some of the NRQED couplings to
next-to-leading order in $\alpha$. We first summarize the
next-to-leading matchings which are already given in the literature,
followed by a calculation of a final class of matchings which have
not been previously published, but which are required to obtain
the hyperfine structure to the order we require. \S3  gives a quick summary
of NRQED powercounting, as obtained in ref.~\patrick. These
powercounting rules are then used to systematically identify all
possible contributions to the $O(m  \alpha^5)$ hyperfine splitting.
The main result of this section is that the complete $O(m \alpha^5)$
contribution is obtained in NRQED using precisely the same graphs as for the
$O(m  \alpha^4)$ contribution, but with coupling constants which
are matched to relativistic QED to higher order in $\alpha$.  These
results are finally brought together in \S4, where we present our
results for the hyperfine splitting. We do this for both the ground state
and for the excited states of positronium.

\section{NRQED}

We start with the lagrangian density of NRQED as applied to
nonrelativistic electrons and positrons, which is obtained from full
QED by integrating out all virtual physics at scales greater than
the electron mass, $\Lambda \gsim m$.

\subsection{The Lagrangian}

\fig\feynrules

The fields representing
the low energy degrees of freedom are $\psi$, $\chi$ and $A_\mu$,
which respectively represent the nonrelativistic electron, the
nonrelativistic positron and  photons with energy less than $m$.
The lagrangian is: $\Scl = \Scl_{\rm photon} + \Scl_{\rm 2-Fermi} +
\Scl_{\rm 4-Fermi} + \cdots$, with:\foot\redundant{We omit the four
fermion terms proportional to $\ss c_7$ and $\ss c_8$
which are listed in ref.~\patrick, because they are redundant
in the sense that they may be expressed in terms of those we
display.}
\label\nrqedlagr
\eq
\eqalign{
\Scl_{\rm 2-Fermi} &= \psi^\dagger \left[ i D_t +
{\Bfd^2 \over 2m} + { \Bfd^4 \over 8 m^3} + c_1 \; \pmb{$ \sigma$} \cdot \Bfb
\right. \cr
& \qquad + c_2 \; \Bigl( \Bfd \cdot \Bfe - \Bfe \cdot \Bfd \Bigr) + c_3 \;
\pmb {$\sigma$} \cdot \Bigl( \Bfd \times \Bfe - \Bfe \times \Bfd \Bigr)
 + \cdots \Bigr] \psi + (\psi \to \chi) , \cr
\Scl_{\rm 4-Fermi} &= c_4 \; \psi^\dagger (\pmb {$\sigma$} \, \sigma_2 )
\chi^* \cdot \chi^\sst ( \sigma_2 \, \pmb {$\sigma$} ) \psi + c_5
\;\psi^\dagger
(\sigma_2) \chi^* \; \chi^\sst (\sigma_2) \psi \cr
&\qquad \qquad + c_6 \; \Bigl[ \psi^\dagger (\sigma_2 \, \pmb {$\sigma$})
\Bfd^2 \chi
\cdot \chi^\dagger (\sigma_2 \, \pmb {$\sigma$}) \psi \Bigr] + \cdots, \cr
\Scl_{\rm photon} &= \hf \Bigl( \Bfe^2 - \Bfb^2 \Bigr) + c_9 \;
A_0(\bfk) \, {\bfk^4 \over m^2} \, A_0(\bfk) \cr
& \qquad \qquad - c_{10} \;
A_i(\bfk) \, {\bfk^4 \over m^2} \, A_j(\bfk) \; \left(
\delta_{ij} - {k_i k_j \over \bfk^2} \right) + \cdots . \cr}
\eeq
Here $D_t = \partial_t - iqe A_0$ and $\Bfd = \nabla - iqe \Bfa$,
where $q = -1$ for electrons and $q=+1$ for positrons. \pmb{$\sigma$}\
represents the usual Pauli spin matrices. The Feynman rules corresponding
to this lagrangian are listed in Figure \feynrules.

\subsection{Matching at Leading-Order}

\ref\kinoshita{M. Nio and T. Kinoshita, \prd{53}{96}{4909}.}

The various coefficients, $c_i$, appearing in this expression are
calculable functions of $\alpha$ and $m$ which are obtained by
integrating out scales larger than $m$ using QED. They may be
conveniently determined by computing scattering processes for
free electrons and positrons using both full QED and the NRQED
lagrangian, Eqs.~\nrqedlagr. Equating the results to a fixed order
in $\alpha$ and $v$ completely determines the constants $c_i$
to this order.

Performing this operation at tree level in QED gives the
lowest-order results in $\alpha$ \lepage:
\label\treematching
\eq
c^{(0)}_1 = {qe \over 2m} , \qquad
c^{(0)}_2 = {qe \over 8m^2} , \qquad
c^{(0)}_3 = {iqe \over 8m^2} , \qquad
c^{(0)}_4 = - \, {\pi \alpha \over m^2} ,
\eeq
and
\eq
c^{(0)}_5 = c^{(0)}_6 = c^{(0)}_9 = c^{(0)}_{10} = 0.
\eeq

\fig\treematchingonephoton

\midinsert
 \centerline{\epsfxsize=5.5cm \epsfbox  {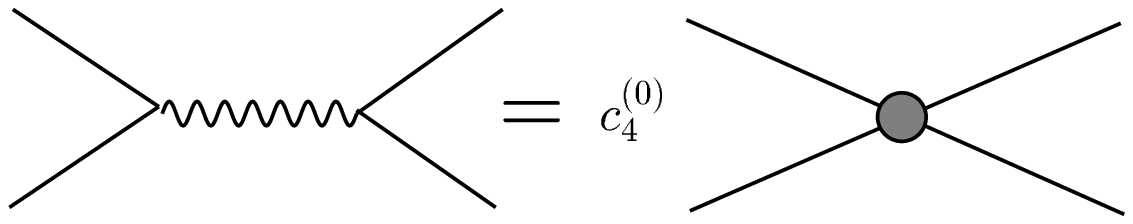}}
\bigskip
\centerline{\bf Figure \treematchingonephoton}
\medskip\noindent
\vbox{\baselineskip=10pt \centerline{The matching which determines
$c^{(0)}_4$.}}
\endinsert

To see how these arise consider, for example, the contribution
to the effective four-fermi couplings $c_4$ and $c_5$.
These four-fermi operators reproduce
in NRQED the effects of the tree-level $s$-channel
annihilation graph of QED, given in Figure \treematchingonephoton.
This must be reproduced by an effective interaction because the
photon which is exchanged must necessarily involve four-momenta
of order $m$, and so cannot appear in the effective theory.

In performing the matching, we evaluate all diagrams at
threshhold --- \ie\ with vanishing external three momentum
in the center of mass frame of the electron-positron  pair.
We use this choice of external momentum purely for convenience,
and the matching could equally well be performed with non-zero
(but nonrelativistic) three-momenta for the electron and positron,
giving precisely the same NRQED coefficients. Notice also that
care must be taken that the scattering amplitude is normalized
in the same way when comparing the QED and NRQED results,
since the covariant normalization of relativistic spinors involves
additional factors of $\sqrt{2E}$ compared to nonrelativistic
applications.

The contributions to the coefficients $c_4$ and $c_5$
are distinguished from one another by separately evaluating the QED
amplitudes for the electron-positron spins combined into a spin triplet
($S=1 \leftrightarrow c_4$) or a spin singlet ($S=0 \leftrightarrow c_5$).
It follows that $c^{(0)}_5$ must vanish since
the spin-singlet combination of the electron and positron
is forbidden by charge conjugation invariance from
receiving a contribution from the $s$-channel
annihilation graph of Fig.~\treematchingonephoton.

Although the $t$-channel photon-exchange graph is also involved in
electron-positron scattering at tree level in QED, it does not contribute
to either $c^{(0)}_4$ or $c^{(0)}_5$. Indeed,
for nonrelativistic electrons and photons
the energy of the exchanged photon is much below the electron mass
and so this scattering is described by the same $t$-channel graph
in NRQED. As a result tree-level $t$-channel photon exchange does
not contribute to any of the NRQED four-fermi operators in
the matching process.

It bears emphasis that the matching calculation does not involve
any bound states at all, since it is done using only the scattering
of free electrons and positrons. Matching is also the only stage
of the calculation which involves QED diagrams. Once the couplings
of the NRQED lagrangian are obtained in this way, only they are
used in the subsequent bound state part of the calculation. This
separation of the matching from the bound state calculations
lies at the heart of NRQED's simplicity. For example, this
separation permits the use of different gauges in the relativistic
and nonrelativistic parts of the calculation, since the gauge
choice when using NRQED is independent of the gauge used in
QED, so long as one computes only gauge-invariant quantities.
This permits the convenience of using a covariant gauge, like
Feynman gauge, in the QED part of the calculation, while keeping
Coulomb gauge for NRQED calculations.

\subsection{Matching at Next-to-Leading Order}

More accurate calculations, such as those of interest in this
paper, require the next-to-leading corrections to
the coefficients of Eqs.~\treematching. Some of these corrections
are already given in the literature, and some we compute
here for the first time. Since our new calculations are for the
coefficients of the four-fermi coupling constants, $c_4$ and
$c_5$, we discuss the corrections to these quantities in some
detail, and give only a brief summary of the higher-order
corrections to both $\Scl_{\rm photon}$ and $\Scl_{\rm 2-Fermi}$.
We write $c_i = c^{(0)}_i + c_i^{(1)} + \dots$ with $c_i^{(0)}$ as
given in the previous section, and now concentrate on
computing the next corrections, $c_i^{(1)}$.

Among the simplest higher-order corrections to the NRQED lagrangian,
Eq.~\nrqedlagr, are the contributions to $\Scl_{\rm photon}$. To lowest
order these are produced by vacuum polarization, which gives:
\label\vacpol
\eq
c^{{(1)}}_9 = c^{{(1)}}_{10} = {\alpha \over 15\pi}.
\eeq

Similarly, QED one-loop vertex corrections modify the couplings
in $\Scl_{\rm 2-Fermi}$, to give \thesis, \kinoshita:
\label\oneloopcorr
\eq
c^{{(1)}}_1 = {qe \over 2m} \; \left( {\alpha \over 2\pi} \right), \qquad
c^{{(1)}}_2 = - \; {qe \over 8m^2} \; \left( {\alpha \over \pi} \right) \left[
\ln \left( {2 \Lambda \over m} \right) - {20 \over 9} \right] , \qquad
c^{{(1)}}_3 = {iqe \over 8m^2} \; \left( {\alpha \over \pi} \right) .
\eeq
Here $\Lambda$ is the ultraviolet cutoff used to regulate NRQED loop graphs
which arise when matching. In any calculation of physical properties,
such divergences cancel amongst themselves, or against
explicit cutoff dependence which arises from divergences
in NRQED loop graphs.

We now turn to the next-to-leading corrections to $\Scl_{\rm 4-Fermi}$.
These corrections may be divided into the following three classes
according to the topology of the one-loop QED graphs which are involved:

\topic{One-Photon Annihilation}
These corrections consist of QED graphs which describe
one-loop corrections to the tree-level process of the
$s$-channel exchange of a single virtual photon. As before
$t$-channel exchange of a single virtual photon does not contribute
corrections to $\Scl_{\rm 4-Fermi}$.
\topic{Two-Photon Annihilation}
These consist of the QED `box' graphs which describe the
$s$-channel exchange of two virtual photons.
\topic{$t$-Channel Two-Photon Exchange}
The final class consists of QED box graphs which describe the
$t$-channel exchange of two virtual photons. Although
$t$-channel one-photon exchange does not contribute to
NRQED four-fermion interactions, $t$-channel two-photon
exchange {\it does} contribute because there is a region of
phase space for which the loop momentum is larger than the
electron mass, and so which does not appear in the corresponding
two-photon exchange graphs of NRQED. The corresponding
physics is therefore put back into the effective field theory
through the NRQED four-fermi interactions.
\endtopic

We now describe, in turn, the matching due to each of these classes
of graphs. While the contributions to $c^{(1)}_4$ and $c^{(1)}_5$
due to the first two of these may be found in the literature
\thesis, \kinoshita, those due to the third class we present here
for the first time.

\fig\loopmatchingonephoton
\midinsert
 \centerline{\epsfxsize=6.5cm \epsfbox  {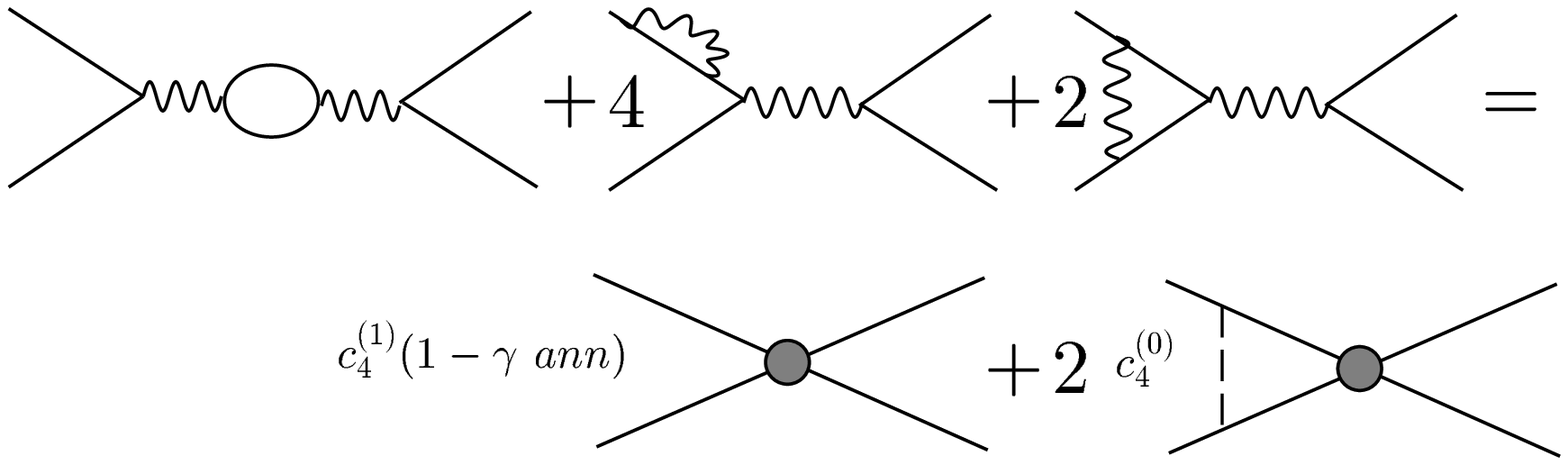}}
\bigskip
\centerline{\bf Figure \loopmatchingonephoton}
\medskip\noindent
\vbox{\baselineskip=10pt \centerline{The graphs which give the
one-photon annihilation contribution to $c_4^{(1)}$.}}
\endinsert

Matching due to one-loop corrections to single-photon
annihilation are described by the graphs of Figure \loopmatchingonephoton.
The contributions to the scattering amplitude of electrons and positrons
at threshhold from the QED graphs on the left-hand-side of the equality
in this figure must be equated to the contributions of the NRQED graphs
on the right-hand-side. Separately evaluating the QED graphs for a
spin-singlet and spin-triplet $e^+ e^-$ state, and solving the resulting
equalities for $c^{(1)}_4$ and $c^{(1)}_5$ then gives \thesis, \kinoshita:
\label\oneloopsinglephoton
\eq
c^{{(1)}}_4 (\hbox{1-$\gamma$ ann}) = {44 \alpha^2 \over 9 m^2} ,
\qquad\qquad
c^{(1)}_5 (\hbox{1-$\gamma$ ann}) = 0.
\eeq
Just as for tree-level matching, the contribution to the spin-singlet coupling,
$c_5$, vanishes by virtue of charge-conjugation invariance.

Although all possible one-loop $s$-channel single-photon
exchange QED graphs appear on the left-hand-side of
Fig.~\loopmatchingonephoton, the same is not true of
the NRQED graphs appearing on the right-hand-side. An
important issue is therefore how to determine which
NRQED graphs must be included to any given order in
the matching process.
The simplest way  to do this is to imagine performing the
matching slightly off-threshold, \ie\ with
the external particles having a small velocity in the
center of mass frame. As mentioned earlier, this does not
affect the value of the NRQED coefficients. Which NRQED
diagram must be kept is then
 decided by counting powers
of $\alpha$ and $v$, with $v \sim \alpha$ kept in mind
for bound-state applications.

For example, for the present purposes of computing the $O(m \alpha^5)$
hyperfine splitting, we show in \S3 that we require both $c_4$
and $c_5$ to next-to-leading order, $O(\alpha^2)$. This implies
both of these couplings must be matched to QED with an accuracy
up to order $\alpha^2 v^0$. The NRQED diagram proportional to
$c_4^{(1)}$ is of order $\alpha^2 v^0$. The loop diagram involving
the exchange of one Coulomb photon contributes to order \foot\onec{This
diagram is proportional to $\ss \alpha^2 \int d^3k /(( \vec p^2 - \vec k^2 +
i \epsilon)(\vec k - \vec p)^2) \simeq \alpha^2/ p$ where $\ss p= mv$
is the external momentum.} $\alpha^2/v$
and cancels a similar term in the QED one loop vertex correction.
 At threshold, the $1/v$ contribution becomes
\foot\irreg{We regulate
all such infrared divergences
by including a photon mass, $\ss \lambda$, into photon propagators.}
 a $1/ \lambda$ infrared divergence
which, cancels a similar term in the QED diagram.
Because the coefficients of the effective theory describes
only high-energy virtual effects in QED, it is a general result that
matching always produces infrared finite values for
them.

Notice also that all other NRQED loop graphs which could appear on the
right-hand-side necessarily involve additional powers of the
electron or positron velocity, $v$, and so give contributions
to $c_4$ which are smaller than $O(\alpha^2)$. For example,
since the Feynman rule for the emission of a soft photon is
proportional to $ep/m = ev$, the loop graph obtained by dressing
a four-fermi interaction with a soft-photon line rather than
a Coulomb line, contributes a term of order $\alpha \, v^2 c_4^{(0)}$
to the scattering amplitude, representing a correction to
$c_4$ which is of order $\alpha^4$ in any bound-state application.
In this way it may be seen that only the Coulomb-exchange
NRQED loop need be kept in Fig.~\loopmatchingonephoton.

\fig\sloopmatching
\midinsert
 \centerline{\epsfxsize=7.5cm \epsfbox  {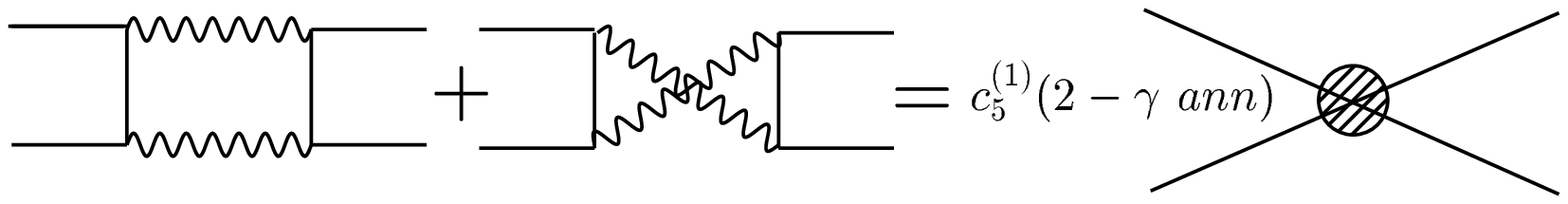}}
\bigskip
\centerline{\bf Figure \sloopmatching}
\medskip\noindent
\vbox{\baselineskip=10pt \centerline{Diagrams which contribute to
two-photon annihilation contributions to $c^{(1)}_5$.}}
\endinsert

We next turn to the QED box graphs describing $s$-channel
electron-positron annihilation into two virtual photons, Figure
\sloopmatching. The matching appropriate for these
graphs gives \thesis, \kinoshita:
\label\oneloopff
\eq
c^{{(1)}}_4 (\hbox{2-$\gamma$ ann}) = 0 , \qquad \qquad
c^{{(1)}}_5 (\hbox{2-$\gamma$ ann}) = \left( {\alpha^2\over m^2} \right)
\Bigl( 2 - 2 \ln 2 + i\pi \Bigr).
\eeq
This time charge-conjugation invariance forbids a contribution
to the spin-triplet operator, $c_4$.

Notice, in this last equation, that the one-loop contribution to
$c_5$, has both a real and imaginary part. This imaginary part
causes the low-energy hamiltonian not to be hermitian.
The resulting loss of unitarity in the time evolution is just
what is required to describe the depletion of electrons and
positrons due to their mutual annihilation into real photons. Since
annihilation is a high-energy
effect, it appears in NRQED as an effective four-fermi operator.
This imaginary part may be used to compute the decay rate for
positronium bound states by calculating the imaginary part it
implies for the bound state energy eigenvalue, $E$. The
decay rate for bound state is then given by the familiar
relation, $\Gamma = -2 ~{\rm Im}(E)$.

\fig\tloopmatching

\midinsert
\centerline{\epsfxsize=9.5cm\epsfbox{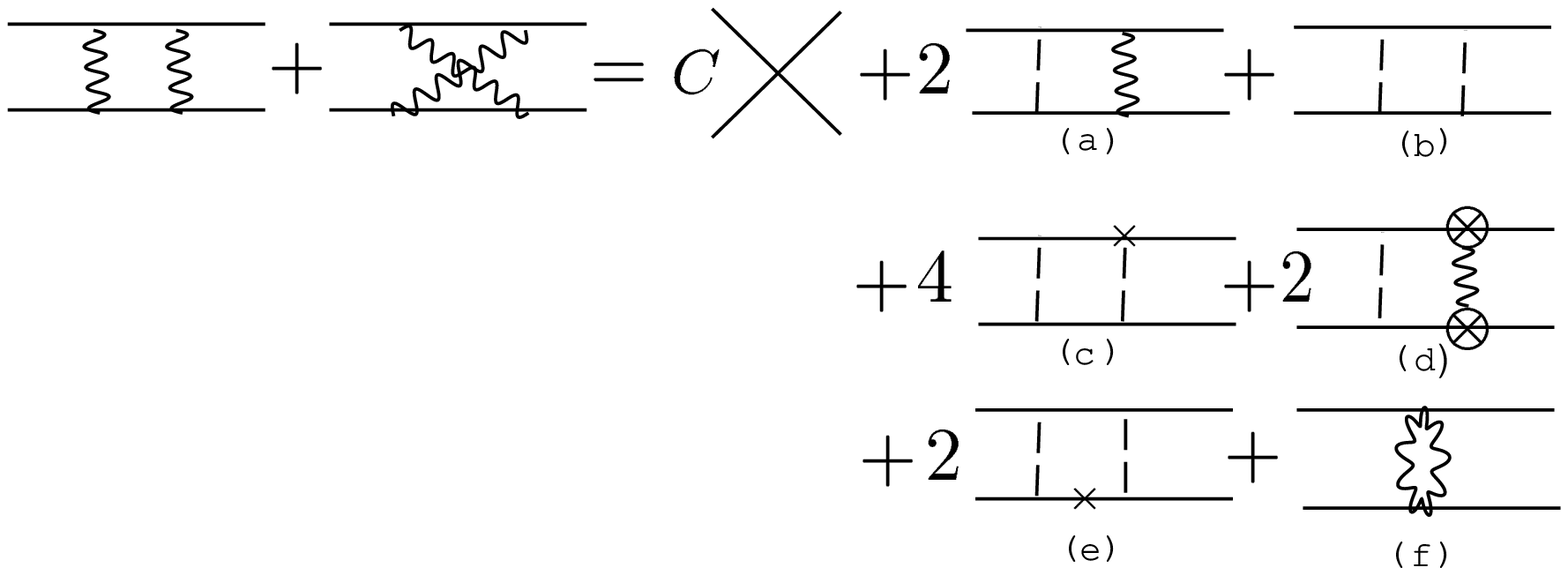
}}
\bigskip
\centerline{\bf Figure \tloopmatching}
\medskip\noindent
\vbox{\baselineskip=10pt \centerline{The one-loop $t$-channel matching
diagrams.}}
\endinsert

We now turn to the contribution to $c_4$ and $c_5$ due to
the $t$-channel two-photon QED exchange graphs, given in Figure \tloopmatching.
Here the diagram proportional to $C$ is meant to represent the
vertex, $c^{(1)}_4$, when the diagrams are evaluated
in a spin 1 state, and $c^{(1)}_5$ when the diagram is
evaluated in a spin 0 state. It is straightforward to verify that
only the given NRQED graphs can contribute to these coefficients
to $O(\alpha^2)$ when $v \sim \alpha$.

In order to determine $c_4^{(1)}$ and $c_5^{(1)}$ separately, we must include
all of the spin-independent NRQED diagrams in the matching. This is
actually more work than is required purely for the purposes of calculating
the hyperfine splitting, since only the difference $c_4^{(1)} -
c_5^{(1)}$ enters into this quantity. Since the spin-independent graphs
cancel in this difference, it suffices to just compute spin-dependent
graphs if one is strictly interested only in the hyperfine splitting.
However, we present here the separate matching for both $c_4$ and
$c_5$, since these two coefficients must be known separately
for other applications, such as for the complete $O(m \alpha^5)$
shift of the positronium energy levels.

Evaluating the left-hand side of Fig.~\tloopmatching\ (the QED graphs)
for a spin-triplet electron-positron configuration gives,
after a straightforward calculation:
\label\QEDSone
\eq
(\hbox{QED})_{S=1} = \alpha^2\; \left[{ -2 \pi m\over \lambda^3}+
{11 \pi\over 12 \lambda m}
+{4\over 3 m^2}+{2\over m^2} \ln \left({\lambda\over m}
\right)\right].
\eeq
Recall $\lambda$ is the infrared-regulating photon mass.

Using the NRQED Feynman rules to calculate the right-hand-side
of Fig.~\tloopmatching\ gives the following contributions.
The only spin-dependent NRQED  diagram, part ($d$) of Figure \tloopmatching,
gives:
\label\figdresult
\eq
\eqalign{
(\hbox{NRQED})_{S=1}(d) &= 2\,\int{d^3 \bfp
\over (2 \pi)^3}\;
\left({-i e \bfp\times \pmb {$\sigma_1$} \over  2m} \right)_i\;
\left({-i e \bfp\times \pmb {$\sigma_2$} \over  2m}\right)_j\;
\left({-1 \over \bfp^2 + \lambda^2} \right)~ \cr
& \qquad \qquad \times \left(\delta_{ij}-
{\bfp_i\bfp_j\over
 \bfp^2+\lambda^2}\right)\; \left({-m\over \bfp^2} \right)\;
\left({-e^2\over \bfp^2+\lambda^2} \right) \cr
&={8 \alpha^2 \over3  m}\;\left( \;\int{dp\;   \bfp^2
 \over (\bfp^2+\lambda^2)^2}\right)\cr
&={2 \pi \alpha^2\over 3 m \lambda} , \cr}
\eeq
where the overall factor of $2$ in the first line
takes into account the two possible ways in which the NRQED diagram
can be drawn.

Similarly, diagram ($f$) of Fig.~\tloopmatching\ gives:
\label\diagfresult
\eq
\eqalign{
(\hbox{NRQED})_{S=1}(f) &=
\int{d^3\bfp\over (2 \pi)^3}
\left({e^2\over 2m} \right) \left({e^2\over 2m} \right)
{1\over \sqrt{\bfp^2+\lambda^2}} \cr
& \qquad \qquad \times \left(
 \delta_{ij}-{\bfp_i\bfp_j\over \bfp^2+\lambda^2}\right)
{-1\over2 \sqrt{\bfp^2+\lambda^2}}{1\over \sqrt{\bfp^2+\lambda^2}}
\left( \delta^{ij}-{\bfp^i\bfp^j\over \bfp^2+\lambda^2}\right) \cr
&={\alpha^2 \over m^2}\left[ { 28 \over 15}-2 \ln 2 +\ln
\left({\lambda \over\Lambda} \right)
\right] .  \cr}
\eeq

All the  other NRQED diagrams can be calculated in a similar manner.
The final result for the sum of the NRQED diagrams, evaluated in a
spin-triplet state is:
\label\nrqedsoneresult
\eq
(\hbox{NRQED})_{S=1} = -2 \, c_4^{(1)}(\hbox{$t$-ch}) +
\alpha^2 \; \left[{ -2 \pi m\over \lambda^3}+
{11 \pi\over 12 \lambda m}
+{28\over 15 m^2}-{2\over m^2}\ln 2  +{2\over m^2}
 \ln\left({\lambda\over \Lambda} \right)\right].
\eeq
Solving for  $c_4(\hbox{$t$-ch})$ then gives the result:
\label\cfourtchan
\eq
c^{{(1)}}_4 (\hbox{$t$-ch}) = - \; \left(
 {\alpha^2 \over m^2} \right) \left[ \ln \left( {\Lambda
\over m } \right) - {4 \over 15} + \ln 2  \right].
\eeq

An identical procedure applies to the $S=0$ state. The QED graphs
are then found to give:
\label\qedszero
\eq
(\hbox{QED})_{S=0} = \alpha^2\; \left[{ -2 \pi m\over \lambda^3}-
{21 \pi\over 12 \lambda m}
+{16\over 3 m^2}+{2\over m^2} \ln \left({\lambda\over m}
\right)\right].
\eeq

For $S=0$, part ($d$) of Fig.~\tloopmatching\ now equals
\label\szerofigdresult
\eq
(\hbox{NRQED})_{S=0}(d) =
 - \, {2 \pi \alpha^2\over   \lambda m}
\eeq
whereas the other NRQED diagrams are left unchanged because they are
spin independent.  The sum of the NRQED diagrams is then found to be
\label\szerosumresult
\eq
(\hbox{NRQED})_{S=0} = -2 \, c_5^{(1)}(\hbox{$t$-ch}) +
\alpha^2 \; \left[{ -2 \pi m\over \lambda^3}-
{21 \pi\over 12 \lambda m}
+{28\over 15 m^2}-{2\over m^2}\ln(2)+{2\over m^2}
 \ln \left({\lambda\over \Lambda} \right)\right].
\eeq

The complete matching result from Fig.~\tloopmatching\
then is
\label\cfivetchan
\eq
c^{{(1)}}_5 (\hbox{$t$-ch}) = - \; \left( {\alpha^2 \over
 m^2} \right) \left[ \ln \left( {\Lambda
\over m} \right) + { 26 \over 15} + \ln 2  \right].
\eeq

The complete one-loop contributions to the coefficients
$c_4$ and $c_5$ are then given by combining the results from
all three classes of graphs:
\label\biganswerforcs
\eq
\eqalign{
c_4^{(1)}&= c_4^{(1)} (\hbox{1-$\gamma$ ann})~+~c_4^{(1)} (\hbox{$t$-ch}) ~=~
{\alpha^2 \over m^2} \left[- \ln \left({\Lambda \over m} \right) +
 { 232 \over 45} - \ln 2 \right]
\cr
c_5^{(1)}&= c_5^{(1)} (\hbox{2-$\gamma$ ann})~+~c_5^{(1)} (\hbox{$t$-ch}) ~=~
{\alpha^2 \over m^2} \left[- \ln \left({\Lambda \over m}
\right) + { 4 \over 15} -
3 \ln 2  + i \pi  \right].}
\eeq

With the NRQED lagrangian now in hand, we next proceed with the
determination of which graphs can contribute to the $O(m  \alpha^5)$
hyperfine splitting in positronium.

\section{Power Counting}

The essence of any effective field theory is its powercounting rules,
since these are what permits the systematic calculation of
observables to any order in small ratios of scales.
Unfortunately, power counting in NRQED is slightly more complicated
than in many effective field theories because of the appearance
of `ultrasoft' photons. Recall that the charged particles in
a QED bound state typically have momenta $p \sim m  \alpha$
and energy $E \sim m  \alpha^2$. Photons
can therefore be emitted with momenta equal to
{\it either} of these scales. Photons having momenta, $k$,
(and energy, $\omega$) of order $m  \alpha$
(which we will refer to as `soft')  do not pose
any problems for powercounting, but those having $k, \omega
\sim m  \alpha^2$ --- the `ultrasoft' ones --- do. Physically,
such ultrasoft photons represent the effects of retardation in
the effective theory.

Fortunately,
these ultrasoft photons have wavelengths which are also large
compared to the size of the bound state, and so their effects
can be organized into slightly more complicated powercounting
rules using what amounts to a multipole expansion in their
couplings to the charged particles. In the final analysis, this
multipole expansion introduces extra suppression by powers
of $\alpha$ into interaction vertices involving ultrasoft photons
and their contribution to the hyperfine splitting starts at order
$m \alpha^6$  \patrick.

\subsection{General Powercounting Rules}

When the dust settles, the powercounting result as applied to positronium
has an appealingly simple form  \patrick. Consider computing
a contribution to a positronium bound-state observable using the
NRQED lagrangian, Eq.~\nrqedlagr, in old-fashioned nonrelativistic
Rayleigh-Schr\"odinger perturbation theory. Each term in this
perturbation series can be expressed graphically, with time
moving to the right, say, and with internal lines denoting the
particles which contribute to the sum over intermediate states.

For NRQED, we use Coulomb gauge, which is the most efficient gauge
for the study of nonrelativistic systems. In this gauge, there are two
types of photon propagator, one corresponding to Coulomb photons
(which are represented by dotted lines in the Feynman diagrams)
and one corresponding to transverse photons (represented by wavy
lines). The Coulomb photons are soft whereas the transverse
photons can be either soft or ultra-soft. Soft transverse
photons are represented by slanted lines in old-fashioned
perturbation theory whereas soft transverse photons are represented
by vertical lines (see  \patrick\ for more details). As mentioned
above, the ultra-soft contributions can be shown to
be  beyond the order we
are interested in this work, and we will therefore consider only
soft transverse photons.
When only soft photons are present -- \ie\ when retardation
effects are neglected -- all NRQED diagrams reduce to a set of
instantaneous interactions, which can  be Fourier transformed
to local interactions in coordinate space separated by
electron-positron propagators.

There are three quantities which determine the size of the contribution
of any such graph to bound-state observables \patrick. For any graph
define $N$ to be the number of electron-positron propagators
separating the instantaneous interactions.~\foot\translation{$\ss N$
 is denoted $\ss N_{TOP}$
in ref.~\patrick.}  The other two quantities are related to the
powers of $\alpha$ and $1/m$ which appear in each of the coupling constants,
$c_i$, of the NRQED lagrangian. Define, then, $\kappa$ and $n$ to be,
respectively,
the total number of powers of $1/m$ (or $v$) and $\alpha$ which
appear in the vertices of the graph of interest.

Suppose $N$, $\kappa$ and $n$ are known for any particular
NRQED graph. Then the contribution of this graph to
the energy-level shift in positronium
depends on $m$ and $\alpha$ (modulo logarithms of
$\alpha$)
through the combination $m  \alpha^p$, where \patrick:
\label\pcrules
\eq
p = 1 + \kappa + n - N.
\eeq
Although the quantity $N$ enters this expression with a negative
sign, inserting additional interactions (and so increasing $N$)
typically involves sufficient additional vertices to ensure that the net
contribution to $p$ increases and that the contribution
from the diagram is therefore suppressed. The only exception to this
statement is repeated insertions of the Coulomb interaction, which
increase both $N$ and $n$ by one but leaves $\kappa$ unchanged
(recall the Coulomb interaction does not contain any powers of inverse
mass). Therefore,  adding any number of Coulomb interactions to a
given diagram  leaves the
value of $p$ unchanged, indicating that one cannot perturb in the Coulomb
interaction,
which must be summed up to all orders. This is accomplished by using
Schr\"odinger wavefunctions for the external lines of the bound state
diagrams and the nonrelativistic Schr\"odinger-Coulomb propagator
for intermediate states.   On the
other hand, it is easy to see that adding any  other interaction
increases the value of $p$ since
they always contain  powers of $1/m$.
There is therefore  only a
finite number of graphs which
 can contribute for any positive choice for $p$.

\subsection{Applications to Hyperfine Splitting}

We may now apply these powercounting arguments to the hyperfine
splitting. An important simplifying feature appears in this specific
application, since the hyperfine splitting compares the energies of two
states having different net spin. As a result it suffices to consider only
graphs for which at least one of the vertices involves a spin-dependent
coupling.

\fig\frthordergraphs

\midinsert
\centerline{\epsfxsize=5.5cm \epsfbox  {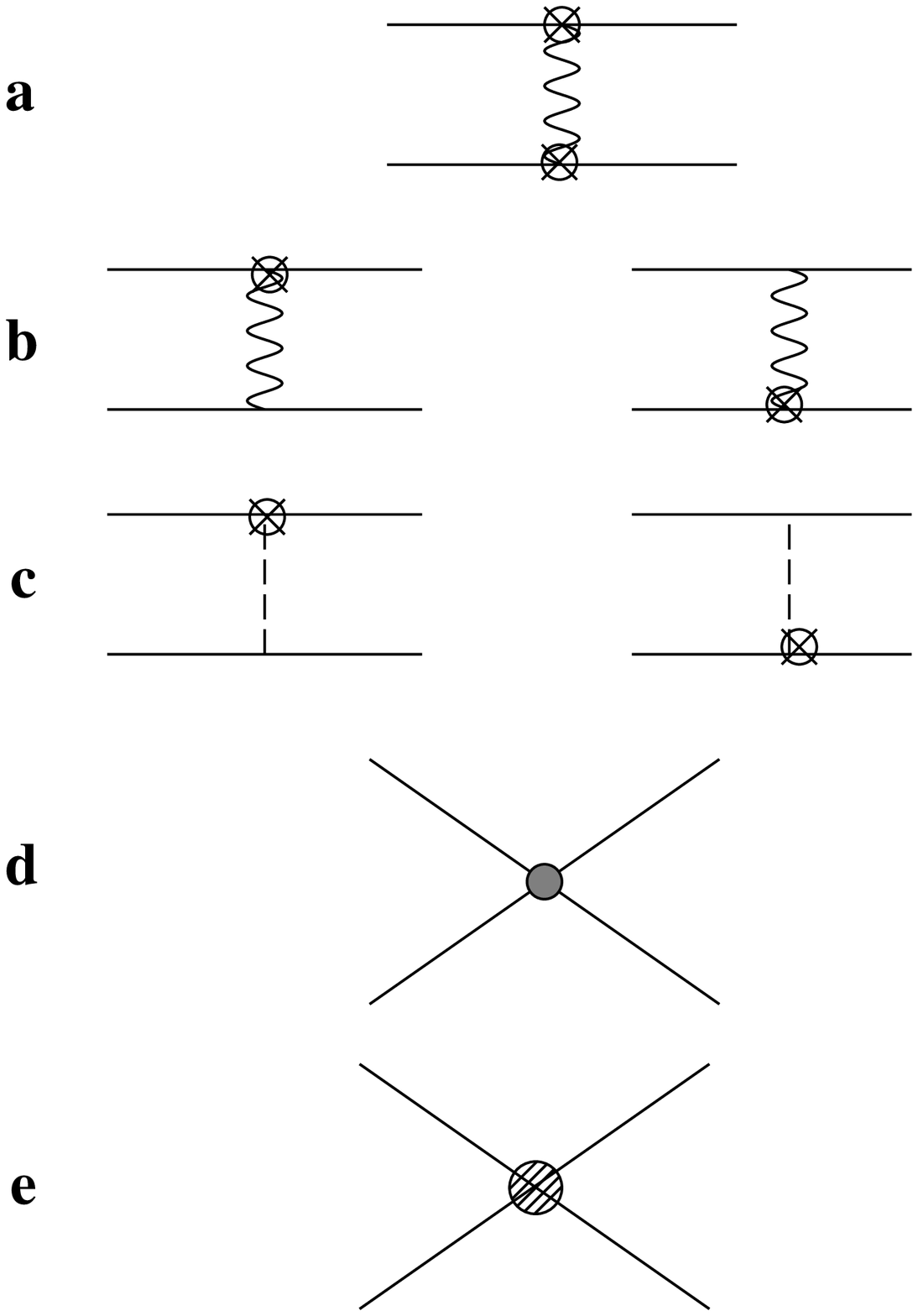}}
\bigskip
\centerline{\bf Figure \frthordergraphs}
\medskip\noindent
\vbox{\baselineskip=10pt \eightrm The NRQED graphs which
contribute to the hyperfine structure at order $\ss m  \alpha^4$.}
\endinsert

To proceed, we start by recapping the powercounting which leads
to the graphs which contribute to the $O(m  \alpha^4)$
hyperfine splitting \scalar. In this case we require $p=4$, so $\kappa+n-N=3$.
Consider first the case $N=0$. In this case we require all possible
graphs containing a single instantaneous photon exchange, subject
to the following two conditions: ($i$) at least one vertex is
spin dependent; and ($ii$) $\kappa + n = 3$. The five
classes of graphs which satisfy these two conditions are displayed
in Figure \frthordergraphs. The coupling which appears
in part (a) of this figure is $c_1$, which appears twice, so inspection of
Eq.~\treematching\ shows that the tree-level contribution,
$c^{(0)}_1$, gives $\kappa = 2$ and $n=1$, as required. The same
is true for parts (b) --- which is linear in $c^{(0)}_1$ --- and
(c) --- which is proportional to $c^{(0)}_3$. Finally, parts (d)
and (e)
involve the four-fermi couplings, $c_4$ and $c_5$, the largest
of which also arises at tree-level, proportional to $\alpha/m^2$, again
giving $\kappa = 2$, $n=1$. It follows that
these graphs contribute to $O(m  \alpha^4)$ provided we use the
leading-order contribution to their couplings.
It is noteworthy that only the graphs of parts
(a), (d) and (e) of Figure \frthordergraphs\ contribute to the
hyperfine splitting of the
ground state, since all of the others vanish when evaluated
in an $s$-wave configuration.

It now remains to consider the case $N > 0$.
Having $N = 1$ implies including one more interaction
in addition to the diagrams shown in Figure (3). It can
be easily verified that any additional interaction (excluding, of
course, the Coulomb interaction) increases the combination
$\kappa +n$ by at least 3.  For example, one can add a transverse photon
coupled to two dipole vertices, introducing the square of the coefficient
$c_1^{(0)}= qe/(2m)$. This increases $\kappa$ by 2 and $n$ by  1. The net
change in $p$ is therefore $3-1=2$ and the diagram
contributes only to order $m \alpha^6$.
 Similarly, adding a Coulomb photon connected to a
Coulomb and  Darwin vertex  increases $\kappa$ by 2 and
$n$ by 1, whereas adding a relativistic kinetic vertex doesn't
change $n$ but increases $\kappa$ by 3. Adding other
interactions necessarily leads to, at best, $O(m \alpha^6)$.

This last argument has immediate implications for
calculating the $O(m  \alpha^5)$ hyperfine structure.
To this order all of the relevant graphs must
still have $N=0$. The next-to-leading result
is therefore obtained from the {\it same} graphs,
Figure \frthordergraphs, but using the next-to-leading
order --- \ie\ $O(\alpha^2)$ --- contributions to
the coefficients, $c_i$. We next show in more detail
how these corrections are computed for the case of the
coefficients of the  four-fermi
interactions, $c_4^{(1)}$ and $c_5^{(1)}$.

\section{Results}

To compute the hyperfine splitting we evaluate the graphs of
Figure \frthordergraphs\ using Coulomb wavefunctions to
describe the initial and final electron-positron lines.
Consider, first, the hyperfine splitting for $s$-wave states.
In this case only graphs ($a$), ($d$) and ($e$) of Fig.~\frthordergraphs\
contribute since the other two graphs contain one vector, \pmb{$\sigma$}, which
can't be dotted into any other vector if $\ell=0$. We find
\label\Epieces
\eq
\eqalign{
\delta E_{n}(a) &= 2 \;\left({ \alpha \over 2 \pi}\right)\int{d^3 \bfp~ d^3
\bfk
\over (2 \pi)^3}\; \Psi^*(\bfp)
\left({-i e (\bfp-\bfk)\times \pmb {$\sigma_1$} \over  2m}\right)_i\;
\left({-i e ((\bfp-\bfk)\times \pmb {$\sigma_2$} \over  2m}\right)_j\;\cr
&\;\;\;\;\;\;\;\;\;\;{-1 \over (\bfp-\bfk)^2}~\left(\delta_{ij}-
{(\bfp-\bfk)_i(\bfp-\bfk)_j\over
 (\bfp-\bfk)^2}\right)\Psi(\bfk)\cr
 &= 2 \;\left({ \alpha \over 2 \pi}\right) \;{ 2 \pi \alpha \over 3 m^2}
<\pmb {$\sigma_1$} \cdot \pmb {$\sigma_2$}>\vert \Psi(0) \vert^2 \cr
 &= 2 \;\left({ \alpha \over 2 \pi}\right) \;
\left( { m \alpha^4\over 6 n^3}
\left[S(S+1)-{3\over 2}\; \right] \delta_{\ell,0} \right)\cr
\delta E_{n}(d) &=  - \; {m^3 \alpha^3 \over 8 \pi n^3} \; \Bigl( c^{(1)}_4 \;
S(S+1)
\Bigr)\;\delta_{\ell,0}
   \cr
&= - \; \left( { m^3\alpha^3 \over 8 \pi n^3}\right) \,\left({\alpha
\over m} \right)^2
\left( -\ln \left({\Lambda \over m}\right) + {232 \over 45} - \ln 2  \right) \;
   \; S(S+1) \; \delta_{\ell,0} \cr
 \delta E_{n}(e) &=  - \; {m^3 \alpha^3 \over 8 \pi n^3} \; \Bigl( c^{(1)}_5 \;
\left[ 2-S(S+1) \right]\;\delta_{\ell,0}
\Bigr)
   \cr
&= - \; \left( { m^3\alpha^3 \over 8 \pi n^3}\right) \,\left({\alpha
\over m} \right)^2
\left(- \ln \left({\Lambda \over m}\right)+ {4 \over 15}- 3 \ln2 + i \pi
\right) \;
  \left[ 2-S(S+1) \right] \; \delta_{\ell,0}\cr}
\eeq
where $n$ and $\ell=0$ are the principal  and orbital quantum number
of the
positronium state of interest, and $S=0$ or 1 is the net intrinsic spin of the
$e^+ e^-$ state.

Using the formula
$\Gamma= - 2 ~{\rm Im}(E)$, we obtain from $\delta E_{n}(e)$
the $O(m \alpha^5)$ decay rate of the $s$-wave state of parapositronium
($S=0$):
\label\decayrate
\eq
\Gamma(n, \alpha^5) = { m \alpha^5 \over 2 n^3} \;
\delta_{\ell,0} .
\eeq
In the rest of the paper, we concentrate on the
hyperfine splitting so we drop the imaginary part
contained in $\delta E_n(e)$.
 Adding the contributions of diagrams ($a$), ($d$) and ($e$) of
Fig.~\frthordergraphs, and taking the difference between $S=1$
and $S=0$ finally gives:
\label\hfresult
\eq
\eqalign{
\Delta E_{\rm hfs} (n,\alpha^5)
 &\equiv \delta E_n(S=1) - \delta E_n(S=0) \cr
&=  - \; \left( {m  \alpha^5 \over 2 \pi n^3} \right)
\left[ \ln 2 + {16 \over 9}   \right]  \delta_{\ell,0}, \cr}
\eeq
in agreement with standard results \theliterature \swave.

The hyperfine splitting  for arbitrary quantum numbers
is computed by modifying the previous calculation in
two ways. First, diagram ($a$) of Fig.~\frthordergraphs\
must be computed for $\ell \neq
0$ states. Notice that
diagrams ($d$) and ($e$) of this figure are nonzero only for
$s$-wave states since they represent contact interactions.
Secondly, diagrams ($b$) and ($c$), which contribute
only to $\ell \neq 0$ states, must be computed.
Since the calculation of these latter
contributions is presented elsewhere \scalar, we
present here just the final result:
\label\genhfresult
\eq
\Delta E_{\rm hfs} (n,\ell,\alpha^5) =  - \; \left( {m  \alpha^5
\over 2 \pi n^3} \right) \left[ \left(\ln 2 + {16 \over 9} \right) \;
\delta_{\ell 0}
- {C_{j\ell} \over 4 (\ell + \hf)} \; \Bigl( 1 - \delta_{\ell 0} \Bigr)
\right],
\eeq
where $j$ is the total angular momentum quantum number,
including electron spin, and the coefficients $C_{j\ell}$ are
given explicitly by:
\label\coeffs
\eq
C_{j\ell}
 = \left\{  \matrix{ {4 \ell + 5 \over 2 ( \ell + 1) ( 2\ell + 3) } \qquad
\hbox{if} \quad j=\ell + 1 \cr
- \; { 1 \over 2 \ell (\ell +1)}  \qquad\hbox{if} \quad j=\ell  \cr
{-4 \ell +1 \over 2 \ell (2 \ell - 1)} \qquad\hbox{if} \quad j=\ell - 1 \cr}
 \right.
\eeq

To summarize, we see that the hyperfine splitting to $O(m  \alpha^5)$ is hardly
more
difficult to obtain in NRQED than is the $O(m  \alpha^4)$ result. The only
extra effort required is obtaining the complete matching of all spin-dependent
effective operators to next-to-leading order in $\alpha$. We have
performed the required $O(\alpha^2)$ matchings for those four-fermi
operators which had not been previously given in the literature. Furthermore,
results for the hyperfine splitting for general $n$ and $\ell$ are obtained
with very little effort.

\bigskip
\centerline{\bf Acknowledgments}
\bigskip

This research was partially funded by the Natural Sciences and
Engineering Research Council of Canada. We thank G. Adkins for bringing
reference \swave\ to our attention.

\listrefs
\vskip -0.5cm
\midinsert
\centerline{\epsfxsize=9.5cm\epsfbox {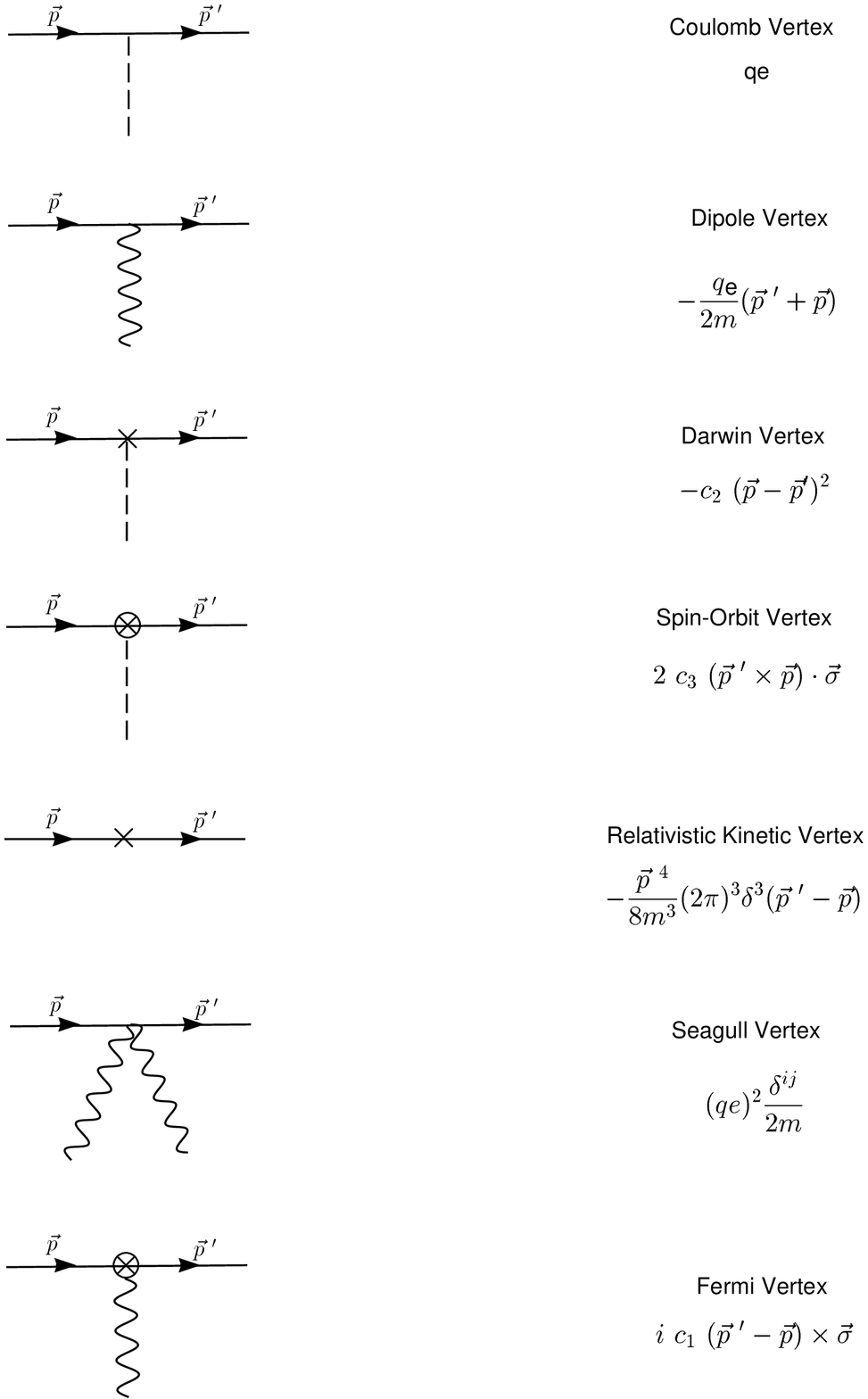}}

\bigskip
\centerline{\bf Figure \feynrules}
\medskip\noindent
\vbox{\baselineskip=10pt \centerline{NRQED Feynman Rules.}
 }
\endinsert
\vskip -0.5cm
\midinsert
\centerline{\epsfxsize=9.5cm\epsfbox {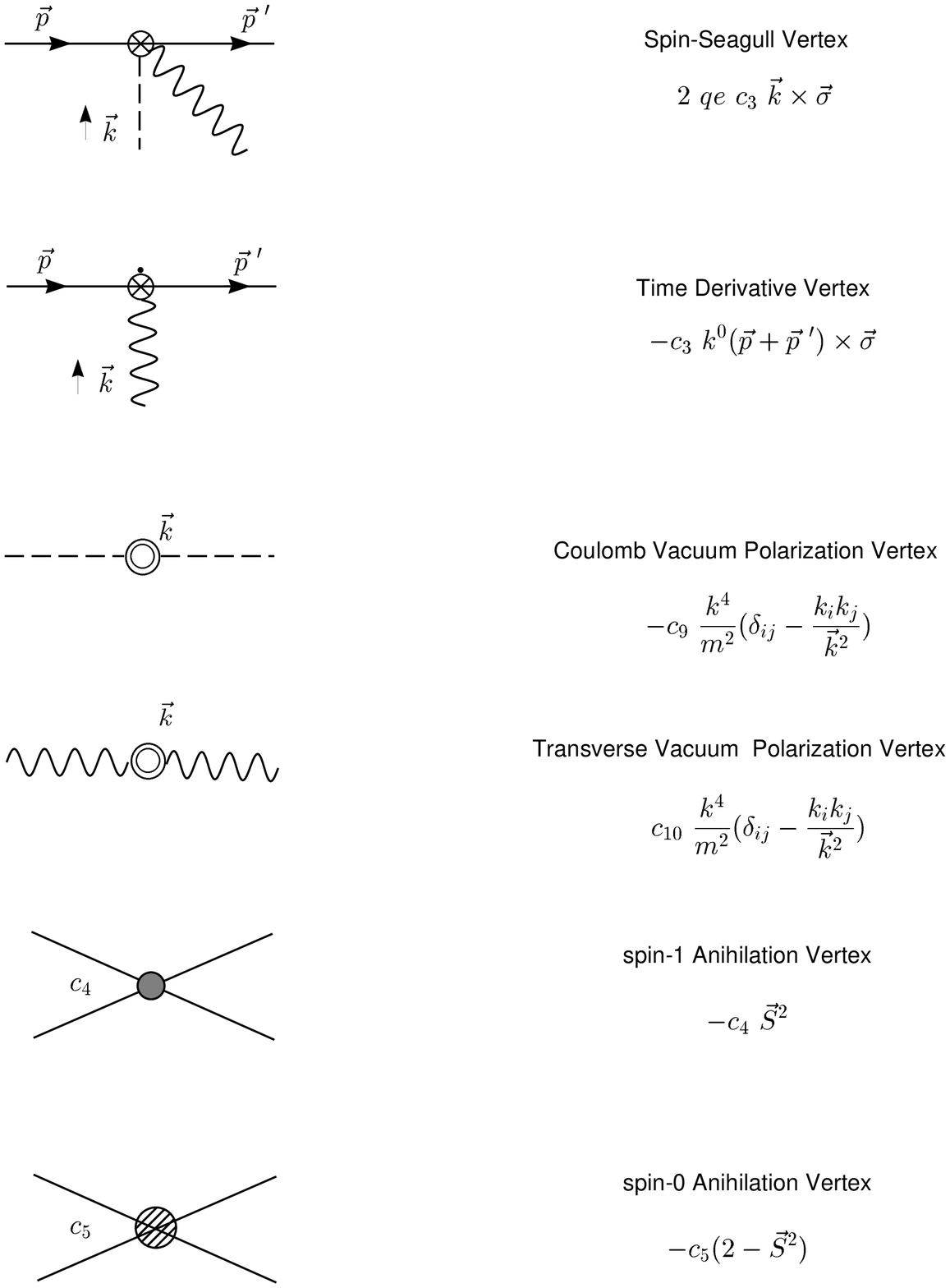}}

\bigskip
\medskip\noindent
\vbox{\baselineskip=10pt \centerline{NRQED Feynman (continued).} .
 }
\endinsert

\bye